# Quantitative model for efficient temporal targeting of tumor cells and neovasculature

M. Kohandel [1,2] [*], C. A. Haselwandter [3,4], M. Kardar [3], S. Sengupta [5], and S. Sivaloganathan [1,2]

[1] Department of Applied Mathematics, University of Waterloo, Waterloo, Ontario N2L 3G1, Canada

[2] Center for Mathematical Medicine, Fields Institute for Research in Mathematical Sciences, Toronto, Ontario M5T 3J1, Canada

[3] Department of Physics, Massachusetts Institute of Technology, Cambridge, Massachusetts 02139, USA

[4] *Present address:* Department of Applied Physics, California Institute of Technology, Pasadena, California 91125, USA

[5] BWH-HST Center for Biomedical Engineering, Department of Medicine, Brigham and Women's Hospital, Cambridge, Massachusetts 02139, USA

**Abstract:** The combination of cytotoxic therapies and anti-angiogenic agents is emerging as a most promising strategy in the treatment of malignant tumors. However, the timing and sequencing of these treatments seem to play essential roles in achieving a synergic outcome. Using a mathematical modeling approach that is grounded on available experimental data, we investigate the spatial and temporal targeting of tumor cells and neovasculature with a nanoscale delivery system. Our model suggests that the experimental success of the nanoscale delivery system depends crucially on the trapping of chemotherapeutic agents within the tumor tissue. The numerical results also indicate that substantial further improvements in the efficiency of the nanoscale delivery system can be achieved through an adjustment of the temporal targeting mechanism.

---

[*] Corresponding author (Email: kohandel@uwaterloo.ca, Tel: 1-519-888-4567, Fax: 1-519-746-4319)

## 1. Introduction

The growth of a tumor beyond an avascular state requires the expansion of its vascular network, a process which is realized through the recruitment of host vasculature (angiogenesis) and/or vasculogenesis. Although the inhibition of tumor angiogenesis represents a promising approach to the treatment and control of cancers, recent preclinical studies have suggested that currently available anti-angiogenic strategies are unlikely to produce significant therapeutic gains on their own, but rather will need to be used in combination with conventional treatments to achieve maximal benefit (Jain 2001, 2005). To date, however, experimental studies combining anti-angiogenic and cytotoxic therapies have shown mixed results (Murata *et al.* 1997, Bello *et al.* 2001, Ma *et al.* 2001, Rofstad *et al.* 2003, Zips *et al.* 2003, Fenton *et al.* 2004, Ma and Waxman 2008), perhaps in part due to differences in scheduling and sequencing of these modalities.

Currently, one major challenge to the successful combination of conventional and anti-angiogenic therapies is that the administration of an anti-angiogenic agent impairs blood flow inside the tumor microenvironment, thus preventing efficient delivery of the chemotherapeutic agent. This difficulty must also be reconciled with the emerging notion of "normalization" of tumor vasculature. The tumor vascular network that arises from abnormal angiogenesis is spatially and temporally heterogeneous with defective endothelium, basement membrane, and pericyte coverage, and is characterized by interstitial hypertension, hypoxia, and acidosis (Jain 2005). Although high global blood flow is a feature of many tumors, the irregular tumor vasculature is very *inefficient* at delivering nutrients, as well as chemotherapeutic drugs, to malignant cells. It has been suggested (Jain 2001, 2005) that the judicious administration of certain anti-angiogenic agents can structurally and functionally "normalize" the abnormal tumor vascular network, rendering the vasculature more conducive to the efficient delivery of both drugs and nutrients to the targeted cancer cells. This transient normalization is characterized by more regular vascular morphology and basement membrane structure, increased pericyte coverage, and decreased hypoxia and interstitial fluid pressure. Recent experimental and clinical studies have shown that blockade of VEGF (vascular endothelial growth factor) signaling, passively prunes some of the immature and leaky vessels of tumors, and actively remodels the remaining vasculature, resulting in a more normalized network (Tong *et al.* 2004, Willett *et al.* 2004, Winkler *et al.* 2004). Even more recently, it has been shown that creation of perivascular nitric oxide gradients may also result in the normalization of tumor vasculature (Kashiwagi *et al.* 2008).

A severe limitation to taking advantage of a normalized vascular network is that such a state lasts for only a short period of time (Winkler *et al.* 2004, Franco *et al.* 2006, Hormigo *et al.* 2007, Batchelor *et al.* 2007). After the transient window of vascular normalization has passed, both tumor oxygenation and penetration of chemotherapeutic drugs decrease. The ensuing hypoxia activates hypoxia-inducible factor (HIF)-1 , up-regulating many genes involved in angiogenesis, and renders tumor cells resistant to chemotherapeutic agents (Pugh *et al.* 2003, Bristow and Hill 2008). Thus, spatial and temporal tumor targeting play a critical role in devising efficient combination therapeutic strategies.

Sengupta *et al.* (2005) recently designed a novel delivery system (termed a nanocell) comprising a nanoscale pegylated-lipid envelope coating a nuclear nanoparticle. A chemotherapeutic agent (doxorubicin) is conjugated to the nanoparticle and an anti-angiogenic agent (combretastatin) is trapped within the lipid envelope. The nanocells extravasate into the tumor through the enhanced permeation and retention (EPR) effect (Yuan *et al.* 1994, Sengupta and Sasisekharan 2007) (see Fig. 1), a consequence of the highly leaky nature of tumor vasculature (having pores with diameters of 400-600nm). This is clearly visible in Figure 1 by the preferential accumulation of nanocells (labeled with quantum dots) in the tumor compared to other vascularised tissues: the nanocells are spatially restricted within normal vasculature but extravasate out from the tumor vasculature (Sengupta et al. 2007). The rapid release of the anti-angiogenic agent results in at least a partial collapse of the network of tumor blood vessels. The entrapped nanoparticles then slowly release the chemotherapeutic drugs, which remain localized due to the disruption of the nearby vasculature. Sengupta *et al.* (2005) compared the effects of sequential drug delivery using nanocells with several conventional approaches on mice with B16:F10 melanomas or Lewis lung carcinomas. Animals treated with nanocells containing both drugs showed a better tumor response than any of the other treatment groups.

While the nanoscale delivery approach outlined above produced a markedly improved effect on tumor control, there remains potential for further refinement of the release kinetics. This motivated us to adapt a mathematical model (Kohandel et al., 2007), which incorporates tumor cells, vascular network, as well as their interplay, and the effects of chemotherapy and anti-angiogenic therapy to the experimental system studied by Sengupta et al. The details of this model are described in Sec. 2; Sec. 3 discusses the mechanism for the synergistic effect of the nanocell treatment suggested by our model, and how it may be possible to improve the efficiency of the nanocell treatment even further.

## 2. Methods

In order to devise an efficient combination of cytotoxic and anti-angiogenic therapies, it is essential to take into consideration the mechanism and timing of tumor vessel response to anti-angiogenic agents, as well as the coupling between tumor growth, the vascular network, and response to cytotoxic agents. We have recently developed a mathematical model that incorporates tumor cells and the vascular network, as well as their interactions, and applied it to study the combination of anti-angiogenic and radiotherapeutic treatments (Kohandel *et al.* 2007). The experimental data of Winkler *et al.* (2004) were used to estimate the model parameters and validate its predictions. The results indicated that application of anti-angiogenic therapy, which temporarily results in better delivery of therapeutic agents, in advance of radiotherapy is the most effective approach, consistent with the experimental results.

In this paper, we build upon a previous mathematical model (Kohandel *et al.* 2007) by including the effects of chemotherapy and the sequential release kinetics of nanocells. In formulating the model our guiding principle is to make minimal assumptions about the underlying phenomenology of cellular processes, while incorporating the essential features of the experiments by Sengupta *et al.* (2005). A key feature of the latter is the release kinetics of the drugs in nanocells: combretastatin has a rapid release (reaching significant levels within 12 h), while doxorubicin releases more slowly extending over 15 days (compared to approximately 4 days for liposomes). Our mathematical model incorporates this temporal targeting profile and allows complete control over other possible factors contributing to the increased effectiveness of the nanocell treatment, such as differences in the total amount of drugs delivered to the cancer cells. Thus, we can directly test and further investigate the temporal targeting mechanism proposed by Sengupta *et al.* (2005).

### 2.1 Tumor cells and vascular network

Following previous studies of tumor growth, we model the density of cancer cells (at position *x* at time *t*) by a spatio-temporal field $n(x,t)$ according to a variant of the Fisher equation (below). The novel aspect of our approach is to similarly introduce a field $m(x,t)$ to model the density of blood vessels. This circumvents accounting for the precise nature of blood flow, and as demonstrated earlier (Kohandel *et al.* 2007), the key features of the interplay between growth and blood supply can be captured the evolution equations

$$\frac{\partial n}{\partial t} = \nabla^2 n + n(1-n) + \tilde{\alpha}_1 m(x,t)n, \qquad (1.1)$$

$$\frac{\partial m}{\partial t} = \tilde{D}_2 \nabla^2 m + m[\tilde{\alpha} + \tilde{\beta}m + \tilde{\gamma}m^2] + \tilde{\alpha}_2 n(x,t)m. \qquad (1.2)$$

For simplicity, the above equations are presented in dimensionless form. They are related to the corresponding dimension-full equations via the transformations $t \to \rho t$, $x \to \sqrt{\rho/D_1}\, x$, and $n \to n/n_{\lim}$ (Kohandel *et al.* 2007), where $\rho$ is the net proliferation rate, $D_1$ is the diffusion coefficient of tumor cells, and $n_{\lim}$ is the "carrying capacity" for tumor cells.

For $\tilde{\alpha}_1 = 0$, Eq. 1.1 is the (dimensionless) Fisher equation, which has two fixed points; an unstable fixed point at $n^* = 0$ (no population at all), and a stable fixed point at $n^* = 1$ (where the population saturates to the carrying capacity). In the absence of the nonlinear term, i.e., for the simple exponential form and $\tilde{\alpha}_1 = 0$, integrating both sides of Eq. 1 leads, for a constant diffusion coefficient, to a simple exponential increase in the number of cells. The growth of a tumor beyond an avascular state (up to a maximum size of about 1-2 mm in diameter) requires the development of a vascular network. The additional term $\tilde{\alpha}_1 mn$ in Eq. 1 indicates that tumor growth is enhanced by the presence of vasculature. Mathematically, this results in a stable fixed point at $n^* = 1 + \tilde{\alpha}_1 m^*$, see below.

Following Kohandel *et al.* (2007) [1], we use Eq. 1.2 to take into account the heterogeneous tumor vasculature. This coarse-grained model, instead of the exact pattern of vessels, produces islands of vascular and non-vascular networks. For $\tilde{\alpha}_2 = 0$, and setting $\tilde{\alpha} = -1$, $\tilde{\beta} = 3$, and $\tilde{\gamma} = -2$, we obtain two stable fixed points for $m^*$ at 0 and 1, and one unstable one at ½. Starting from a random (positively distributed and close to zero) initial configuration for $m(x,0)$, Eq. 1.2 produces randomly distributed islands of $m = 1$ (vascular) and $m = 0$ (non vascular). The last term in Eq. 1.2, $\tilde{\alpha}_2 nm$, represents the effect of tumor cells on the development of vessels. We assume that the tumor cells produce the proangiogenic cytokines, leading to the extension of the vascular network; in our phenomenological approach, we assume that the higher density of cancer cells creates higher vascular density. In fact, a non-zero $\tilde{\alpha}_2$ shifts the stable fixed points to

$$m^* = \frac{1}{4}(3 + \tilde{\alpha}_1\tilde{\alpha}_2 \pm \sqrt{8(\tilde{\alpha}_2 - 1) + (3 + \tilde{\alpha}_1\tilde{\alpha}_2)^2}).$$

For example, for $\tilde{\alpha}_1 = 1.1$ and $\tilde{\alpha}_2 = 0.9$ (obtained from fits to experimental data, see the results and discussion section), $m^* = 0.02, 1.97$. Thus the fixed point at 1 moves to a higher value, indicating that the tumor vascular density is higher than in the corresponding normal tissue. This increased value will also be utilized in modeling of the poor delivery characteristics of the tumor vasculature, as well as the effect of anti-angiogenic therapy, see below.

---

[1] Equations (6) and (8) of Kohandel *et al.* (2007) contain typos in the interaction terms $\alpha_1 K(x,t)c$ and $\tilde{\alpha}_1 K(x,t)c$ which, in the notation of Kohandel *et al.* (2007), should in fact read $\alpha_1 m(x,t)c$ and $\tilde{\alpha}_1 m(x,t)c$, respectively.

**2.2 Delivery of nanocells and liposomes**

Following the same strategy, we model the spatio-temporal variations of liposomes and nanocells by concentration fields $C_i(x,t)$, the discrete index $i$ labeling the drug administration at time $t_i$. The permeation of nanoparticles within a tumor depends on their sizes; large nanoparticles of the order of 100 nm (which is the case in Sengupta et al. 2007 experiments) appear to stay close to the vasculature (Perrault *et al.* 2009). Hence, we assume that diffusion of liposomes and nanocells within the tissue surrounding the tumor, as well as re-absorption of these particles into the blood vessels, can be neglected. The evolution of the field $C_i(x,t)$ is then modeled by the dimensionless equation,

$$\frac{\partial C_i}{\partial t} = \widetilde{\delta} \, \Gamma_i(t) m(x,t) \exp\left[-\left(\frac{m(x,t)}{m_{\lim}}\right)^2\right]. \qquad (1.3)$$

The function $\Gamma_i(t)$ represents the (average) concentration of the liposomes and nanocells in the blood vessels (see Sec. 2.4). The tumor vasculature is structurally and functionally abnormal, and the vessels are very inefficient at delivering nutrients and chemotherapeutic drugs. This poor delivery could be due to defective vascular structure, lack of perfusion of tumour blood vessels, inconsistent flow, and elevated interstitial fluid pressure (Jain 2008). However, there is growing evidence that vascular efficiency can be improved with anti-angiogenic therapy through the "normalization" process (Jain 2001, 2005). In Eq. 1.3, the poor delivery of tumor vessels is modeled by the function $m \exp[-(m/m_{\lim})^2]$. For $m_{\lim} = \sqrt{2}$, this function has a maximum at $m=1$, corresponding to the efficient delivery of normal vessels. For $m>1$, which corresponds to tumour vasculature (for example, $\widetilde{\alpha}_1 = 1.1$ and $\widetilde{\alpha}_2 = 0.9$ gives $m^* = 1.97$, as mentioned in the previous section), the delivery decreases. Finally, $m<1$ corresponds to immature or degraded vessels, resulting again in inefficient nutrient or drug delivery. Hence, for a tumor vasculature, decreasing the field $m$ to values close to one, by the administration of an anti-angiogenic agent, results in improved vascular efficiency and better delivery of nutrients and chemotherapeutic agents (Kohandel *et al.* 2007). The exponential term therefore accounts for the poor delivery of vasculature as well as the increase in the delivery of liposomes and nanocells to tumor cells through normalization. One should note that strong dosage of the anti-angiogenic drug may lead to values of $m$ less than one, leading to either poor delivery through immature vessels or complete regression of the vasculature (Jain 2008).

**2.3 Drug release from nanocells and liposomes**

Next, we denote by $c(x,t)$ and $d(x,t)$ the concentrations of free anti-angiogenic and chemotherapeutic agents released from liposomes and nanocells into the tumor tissue, respectively. The temporal and spatial evolution of these fields is modeled by the dimensionless equations

$$\frac{\partial c}{\partial t} = \widetilde{D}_3 \nabla^2 c + \widetilde{\lambda}^{(C)} R_j^{(C)}(x,t) - \widetilde{v}^{(C)} c, \qquad (1.4)$$

$$\frac{\partial d}{\partial t} = \widetilde{D}_4 \nabla^2 d + \widetilde{\lambda}^{(D)} R_j^{(D)}(x,t) - \widetilde{\mu} \, m(x,t) d - \widetilde{v}^{(D)} d. \qquad (1.5)$$

We shall contrast the four types of treatment tested in the experiments via the index $j$=1-4, denoting chemotherapy (NC[D], $j$=1), anti-angiogenic therapy (L[C], $j$=2), simple liposome encapsulating both (L[CD], $j$=3), and nanocells (NC[CD], $j$=4). Free anti-angiogenesis and chemotherapy agents released from nanocells and liposomes are small enough to diffuse through the tumor tissue [first terms in the right side of Eqs. (1.4) and (1.5); $\widetilde{D}_3$ and $\widetilde{D}_4$ are dimensionless diffusion coefficients]. The term $-\widetilde{\mu}\, m(x,t)d$ describes the re-absorption of free chemotherapy drugs into the blood vessels. We assume that no such term is present in the equation for $c(x,t)$ since anti-angiogenic drugs act on normal as well as on abnormal blood vessels, which prevents absorption. However, both equations involve terms of the form $-\widetilde{v}^{(C)}c$ and $-\widetilde{v}^{(D)}d$, which describe the natural decay of free drugs. The release of free chemotherapy and anti-angiogenesis agents in Eqs. (1.4) and (1.5), i.e., $\widetilde{\lambda}^{(C,D)}R_j^{(C,D)}$, proceeds according to

$$R_j^{(C,D)}(x,t) = \sum_i RP_{i,j}^{(C,D)}(t) C_i(x,t), \qquad (2)$$

in which the sum runs over all administration times $t_i$. The release profiles $RP_{i,j}^{(C,D)}(t)$ are either identical to zero or satisfy

$$\sum_i \int_0^{t_f} RP_{i,j}^{(C,D)}(t)dt = 1,$$

where, as above, the sum runs over all administration times and $t_f = 17$ days is the longest time considered in the experiments by Sengupta *et al.* (2005). The above condition on the release profiles fixes the total amount of drugs released from liposomes and nanocells over the time interval considered in the experiments by Sengupta *et al.* (2005) and, thus, ensures a fair comparison of different therapeutic strategies in our model.

Finally, the effects of chemotherapy and anti-angiogenic therapy on cancer cells and blood vessels are modeled by

$$\left.\frac{\partial n(x,t)}{\partial t}\right|_{chemo} = -\widetilde{A}^{(D)}d(x,t)n, \quad \text{for} \quad j=1,3,4,$$
$$\left.\frac{\partial m(x,t)}{\partial t}\right|_{anti} = -\widetilde{A}^{(C)}c(x,t)m, \quad \text{for} \quad j=2,3,4, \qquad (3)$$

where the vertical lines indicate that the above terms are added to Eqs. (1.1) and (1.2), respectively. Here, $\widetilde{A}^{(D)}$ and $\widetilde{A}^{(C)}$ present the strength of chemotherapy and anti-angiogenic therapy, respectively; $d(x,t)$ and $c(x,t)$ are defined by Eqs. (1.4) and (1.5). The details of the above model and its parameterization are discussed below.

### 2.4 Release profiles

To determine the release profiles, $RP_{i,j}^{(C,D)}(t)$ for $j=1-4$, we first note that NC[D] only involves chemotherapy (but does not contain anti-angiogenic therapy), thus $RP_{i,1}^{(C)}(t) = 0$; recall that NC[D] is denoted by $j=1$. Similarly, $RP_{i,2}^{(D)}(t) = 0$ since L[C] ($j=2$) only involves anti-angiogenic therapy. The remaining release profiles should ideally be fixed from *in vivo* release experiments. We expect that the release profiles of combretastatin and doxorubicin in L[CD] are similar to the release profile of combretastatin in L[C], i.e., $RP_{i,3}^{(C)}(t) = RP_{i,3}^{(D)}(t) = RP_{i,2}^{(C)}(t)$ - this is due to the fact that for all these cases, drugs are included inside a liposome. Similarly, since all nanocells (independently of whether they contain combretastatin or not), have liposome on the outer layer, we have $RP_{i,1}^{(D)}(t) = RP_{i,4}^{(D)}(t)$. Thus, we need to determine the functions $RP_{i,1}^{(D)}(t)$, $RP_{i,2}^{(C)}(t)$, and $RP_{i,4}^{(C)}(t)$. The *in vitro* studies in Sengupta *et al.* (2005) show that the release of doxorubicin from NC[D] is delayed relative to the release of combretastatin from the liposome, with an extending time of approximately 4 days for the liposome and an extending time of approximately 15 days for the core of the nanocell. On that basis we take

$$RP_{i,1}^{(D)}(t) = \frac{1}{N_1}\theta(t-t_i)(t-t_i)^{p_{NC}}\exp(-(t-t_i)/\tau_{NC}),$$
$$RP_{i,2}^{(C)}(t) = \frac{1}{N_2}\theta(t-t_i)(t-t_i)^{p_L}\exp(-(t-t_i)/\tau_L), \qquad (4)$$
$$RP_{i,4}^{(C)}(t) = \frac{1}{N_4}\theta(t-t_i)(t-t_i)^{p'_L}\exp(-(t-t_i)/\tau'_L),$$

where $t_i = $ 8, 10, 12, 14, 16 days are the administration times used in the experiments by Sengupta *et al.* (2005), the constants $N_{1,2,4}$ are determined from the normalization condition, see below Eq. (2), and $\theta(x)$ is the unit step function (defined by $\theta(x) = 1$ for $x \geq 1$ and $\theta(x) = 0$ otherwise). Assuming that the release profile of L[C] and the liposome coating of the nanocell are similar, we can set $p_L = p'_L$ and $\tau_L = \tau'_L$, in which case Eq. (2) implies $N_2 = N_4$. Based on Sengupta *et al.* (2005) we take $p_{NC} = 0.3$ and $\tau_{NC} = 15$ days, and $p_L = 0.1$ and $\tau_L = 2$ days. For our modified nanocell therapy (see Sec. 3) we use the same treatment schedules as for NC[CD] but take $p_{NC} = 0.8$ in Eq. (4) with an appropriate normalization factor determined from Eq. (2). This ensures that for all combined therapies the nanocells and liposomes release the same total amount of combretastatin and doxorubicin.

The function $\Gamma_i(t)$ in Eq. (1.3) describes the (average) concentration of the liposomes and nanocells in the blood vessels. In using the same function $\Gamma_i(t)$ for liposomes and nanocells we assume that, once a therapy has been administered, changes to the drug concentration in the blood vessels only result from some natural decay (e.g., adsorption) independent of the chemical composition of the particles. Thus, we set

$$\Gamma_i(t) = \theta(t-t_i)\exp(-(t-t_i)/\tau_D), \qquad (5)$$

where, as above, $t_i$ are again the administration times. There is some indication (Sengupta *et al.* (2005), see also Fig. 1) that $\Gamma_i(t)$ is different for liposomes and nanocells, with an increased delivery of nanocells into the tissue surrounding the tumor through a mechanism other than vascular normalization. This would further increase the effect of NC[CD] relative to L[CD] through an increase in the amount of drugs delivered to the tumor. However, in order to allow a direct investigation of the temporal targeting mechanism proposed by Sengupta *et al.* (2005) we take the $\Gamma_i(t)$ for conventional therapies to be the same as for the nanocell treatment and set $\tau_D = \tau_{NC}$.

**2.5 Parameterization**

Free anti-angiogenesis and chemotherapy agents are small particles which can be assumed to have diffusion characteristics similar to nutrients such as oxygen, for which experimental data is readily available. Thus, we set $\tilde{D}_3 = \tilde{D}_4$ in Eqs. (1.4) and (1.5) and use a value of $\tilde{D}_3$ similar to the diffusion constant of free oxygen in Kohandel *et al.* (2007). As mentioned earlier, the field $m(x,t)$ stands for the average distribution of blood vessels (rather than the exact pattern of vasculature), thus the diffusion coefficients for drugs (or nutrients) are not of the same order as diffusion from a single blood vessel. Moreover, because chemotherapy agents released from the nanocells are transported away through blood vessels in the same way for NC[D] and NC[CD], and $\tilde{D}_3 = \tilde{D}_4$, we use the same re-absorption rate $\tilde{\mu}$ for all treatments. Similarly as in Kohandel *et al.* (2007) for free oxygen, we assume that the natural decay of $c$ and $d$ is not too strong and, hence, that the decay constants $\tilde{\nu}^{(C)}$ and $\tilde{\nu}^{(D)}$ take smaller numerical values than $\tilde{\mu}$. Furthermore, combretastatin released from the liposome decays faster than doxorubicin released from the nanocells (Sengupta *et al.* 2005) and, thus, we take $\tilde{\nu}^{(C)} > \tilde{\nu}^{(D)}$. On the other hand, one may expect that doxorubicin released from the liposome during treatment with L[CD] decays faster than doxorubicin released from the nanocells. This would further decrease the efficiency of L[CD] relative to NC[CD], but to avoid a proliferation of parameters we use the same value $\tilde{\nu}^{(D)}$ for all treatments.

Table 1: Numerical values for the parameters in Eqs. (1) and (3) used in our simulations. In addition to the values shown, we use in the dimensionful equations $D_1 = 0.32$ mm$^2$/day for lung cancer and $D_1 = 0.46$ mm$^2$/day for melanoma, $\rho = 0.35$ day$^{-1}$ for both melanoma and lung cancer (see Sec. 3.1), and a threshold for the detectability of tumor cells $c_{th} = 0.09$ (Kohandel *et al.* 2007). As explained in the main text we have $\tilde{D}_3 = \tilde{D}_4$ and $\tilde{\lambda}^{(C)} = 100\tilde{\lambda}^{(D)}$. The effectiveness of chemotherapy and anti-angiogenic therapy are parameterized by $\tilde{A}^{(D)} = 1.65$ and $\tilde{A}^{(C)} = 0.1$ for lung cancer, and by $\tilde{A}^{(D)} = 3$ and $\tilde{A}^{(C)} = 0.3$ for melanoma (see Sec. 3.1).

| $\tilde{\alpha}_1$ | $\tilde{D}_2$ | $\tilde{\alpha}$ | $\tilde{\beta}$ | $\tilde{\gamma}$ | $\tilde{\alpha}_2$ | $\tilde{\delta}$ | $m_{\lim}$ | $\tilde{D}_3$ | $\tilde{\lambda}^{(C)}$ | $\tilde{\nu}^{(C)}$ | $\tilde{\nu}^{(D)}$ | $\tilde{\mu}$ |
|---|---|---|---|---|---|---|---|---|---|---|---|---|
| 1.1 | 0.005 | −1 | 3 | −2 | 0.9 | 0.8 | $\sqrt{2}$ | 0.02 | 1300 | 3 | 0.1 | 8 |

The amount of drugs administered can be included in our model through the values of the coefficients $\tilde{\lambda}^{(C)}$ and $\tilde{\lambda}^{(D)}$. In the experiments described in Sengupta *et al.* (2005) approximately one hundred times more combretastatin than doxorubicin was injected and, hence, we take $\tilde{\lambda}^{(C)} = 100\tilde{\lambda}^{(D)}$. Moreover, the constants $\tilde{A}^{(C)}$ and $\tilde{A}^{(D)}$ determine the effectiveness of a given therapy and are therefore crucial parameters in our model. To allow a quantitative comparison between the effects of different combined therapies we use the same values $\tilde{A}^{(C)}$ and $\tilde{A}^{(D)}$ for single and combined therapies. It is thereby assumed that the effect of a given therapeutic agent is not influenced by the presence or absence of another therapeutic agent. The values of all parameters appearing in our model are fixed by fitting volume curves for V, L[C], and NC[D] to the corresponding experimental results (Sengupta *et al.* 2005). Table 1 summarizes the parameter values used in our simulations (see Appendix for the sensitivity analysis of parameters). The results of the combined treatments are then predicted by our model without any further assumptions.

While we consider a variety of different interactions in our model, with the aim of not excluding any possible mechanism for the success of the nanocell treatment *a priori*, we find that only the three parameters $D_1$, $\tilde{A}^{(C)}$, and $\tilde{A}^{(D)}$ need to be adjusted to distinguish between lung carcinoma and melanoma. Moreover, we also find that for both lung carcinoma and melanoma the success of the nanocell treatment relies crucially on the temporal release profiles used in Sengupta *et al.* (2005), and on the possibility of re-absorption of chemotherapeutic drugs into the bloodstream (see Sec. 3), which is parameterized through $\tilde{\mu}$. As in Kohandel *et al.* (2007) we use a normalized Gaussian initial condition for *n(x,t)* with variance $\sigma = 0.35$, and a random initial condition for *m(x,t)* evenly distributed between zero and one. Zero initial conditions are considered for the concentrations of the drugs. All simulations are performed on a cubic grid with $50 \times 50 \times 50$ points and no-flux boundary conditions.

### 3. Results and Discussion

#### 3.1 Consistency of model results

To confirm our model, numerical simulations are performed according to the experimental protocol of Sengupta *et al.* (2005) on lung cancer and melanoma (see Fig. 2). In these experiments, $2.5 \times 10^5$ Lewis lung carcinoma cells or $3 \times 10^5$ GFP-positive BL6/F10 melanoma cells were implanted in male C57/BL6 mice, and treatments started when tumors reached 50 mm$^3$ in volume (after about 8 days). The kinetics of tumor growth and blood vessel formation, as well as the data points for the control group (V, red), are used to estimate the related model parameters for the case of lung cancer (see Table 1). The experimental treatment schedules, as well as pharmacokinetics and pharmacodynamics of agents, are used in the simulations to fit the data for single administration of anti-angiogenic therapy (combretastatin-encapsulated liposomes, L[C], brown) and chemotherapy (nanocells containing doxorubicin but lacking combretastatin, NC[D], blue). The corresponding curves for melanoma are obtained by modifying three model parameters describing the diffusion of cancer cells and the effect of therapeutic agents on the vascular network and cancer cells (see Sec. 2). We then perform numerical simulations with the estimated parameters to predict the results for the conventional (a liposome encapsulating both doxorubicin and combretastatin, L[CD], green) and nanocell (NC[CD], purple) approaches for the combination of anti-angiogenic and cytotoxic treatments.

As shown in Fig. 2, the numerical results (solid lines) obtained from our phenomenological model reproduce the major trends observed in the experiments by Sengupta *et al.* (2005) (points) for lung cancer and melanoma. While the conventional combination of combretastatin and doxorubicin only produces a doubling effect compared to the single administration of combretastatin or doxorubicin, nanocells clearly show a more pronounced effect with the same amount of drugs administered. Our simulations strongly support the hypothesis that the increased effect of the nanocell treatment is mainly due to the temporal release achieved through nanocells (Sengupta *et al.* 2005), which is from a mathematical perspective the only feature distinguishing L[CD] and NC[CD] in our model (see Sec. 2)[2]. This is discussed further below.

#### 3.2 Synergistic effect of nanocell treatment

Our model predicts the nanocell treatment can produce a combined effect that is *greater than the sum of its parts* through the two different mechanisms illustrated in Fig. 3. First, the transient normalization resulting from anti-angiogenic therapy enhances the delivery of additional nanocells to the tumor tissue. This in turn increases the effect of the subsequent release of chemotherapy. Second, anti-angiogenic therapy can also cut off the blood supply to the tumor, which effectively traps the cytotoxic drugs within the tumor tissue. Thus smaller amounts of free chemotherapy agents are transported away from the tumor tissue. According to our simulations, the latter mechanism is crucial for obtaining the superior results of NC[CD] reported by Sengupta *et al.* (2005), while normalization only brings a small increase in the effectiveness of NC[CD] relative to L[CD]. However, as discussed below, normalization can still be employed to further improve the efficacy of the nanocell treatment.

---

[2] In Sengupta *et al.* (2005), the results of the single therapies NC[D] and L[C] were also compared to the co-administration of NC[D] and L[C] (NC[D]+L[C]). However, NC[D]+L[C] only showed a negligible effect over either of the treatments alone. This is probably due to a limitation on the total number of nanocells or liposomes which can be taken up by the tumor tissue at any given time. This limitation does not affect any of the other more effective combination strategies considered in Sengupta *et al.* (2005) and, thus, we do not discuss this case in our modeling approach.

As mentioned earlier, our modelling hypothesis is that by adjusting the interplay between normalization and the temporal targeting described in Sengupta *et al.* (2005) an improved therapeutic outcome can be achieved. To validate this hypothesis, we modeled different release profiles for the nanocell core which delay the secretion of chemotherapy relative to anti-angiogenic therapy even further than the release profile considered in the experiments of Sengupta *et al.* (2005) (see Sec. 3). We were careful to adjust the delayed release profiles such that over the course of the treatment the same total amount of anti-angiogenic and chemotherapeutic agents was released as in the combination therapies, L[CD] and NC[CD], considered before. Thus, compared to the conventional nanocell therapy, with our modified release profiles more chemotherapeutic agents are released by nanocells administered at early times, but correspondingly fewer chemotherapeutic drugs are released by nanocells administered at later times.

The simulation results indicate that with the adjusted release kinetics a substantial improvement in the efficacy of the nanocell treatment can be achieved (see the black curves in Fig. 2). As illustrated in Fig. 3, this improved result is due to a combination of normalization, which is mainly effective in the early stages of the therapy and increases the fraction of nanocells delivered to the tumor tissue at early times, and the vascular shutdown induced by anti-angiogenic therapy, which decreases the fraction of nanocells delivered at later times and traps the chemotherapeutic agents within the tumor tissue. Our modified release profiles take advantage of these two effects and lead to better coordination between the arrival of nanocells in the tumor tissue and the release of anti-angiogenic and chemotherapeutic agents. Thus, our simulations suggest that by means of a judicious timing of normalization and the trapping of chemotherapeutic agents in the tumor tissue through the release of combretastatin, a better therapeutic outcome can be achieved.

In summary, mathematical models can be used to simulate various therapeutic scenarios and aid in hypothesis testing "in silico," and conversely to guide experimental research. Following the very successful experimental results of Sengupta *et al.* (2005), and experimental and clinical studies on the normalization process of tumor vasculature (Tong *et al.* 2004, Willett *et al.* 2004, Winkler *et al.* 2004), we have developed a mathematical formulation that captures the qualitative picture proposed by Sengupta and co-workers. Further the model has been used to test our hypothesis for the effective combination of anti-angiogenic therapy and chemotherapy. On the basis of this model we find that the dramatically improved therapeutic effect of the nanocell treatment demonstrated by Sengupta *et al.* (2005) primarily results from the trapping of chemotherapeutic agents, rather than an increase in the number of nanocells delivered to the tumor through normalization. Moreover, we find that through an adjustment of the release kinetics for chemotherapy it may be possible to substantially improve the efficacy of the nanocell treatment. As a result of our promising computational results, it seems clear that more experimental and preclinical data are required to further validate this strategy for improved therapeutic outcome. In particular, our model results suggest that nanocells with a longer delay in the secretion of chemotherapy relative to anti-angiogenic therapy (compared to the release profile considered in the experiments of Sengupta *et al.* (2005)) may further increase the efficiency of the treatment. Our computational approach could be also used to design nanocells for other types of cancers, and to quantitatively determine promising release profiles.

**Appendix**

In this appendix, we study the effects of small variations in the values of the parameters on the model outcome. To study parameter sensitivity, we calculate the effect of small changes of the parameters on final tumor volume (at day 17). We determine the percent change in the final tumor volume when the model parameters are perturbed from their estimated values by 5% (de Pillis *et al.* 2005). Fig. A1 shows the results for the case of combination of therapies using nanocells and liposomes, i.e. NC[CD] and L[CD], in lung cancer. One should note that changing $\tilde{\lambda}^{(C)} = 100\tilde{\lambda}^{(D)}$ has the same effect as varying $\tilde{\delta}$; in addition, the parameters $\tilde{\alpha}$, $\tilde{\beta}$, and $\tilde{\gamma}$ are not changed as they are responsible for the creation of the vascular network (see main text). The results indicate that the parameters $\tilde{\alpha}_1$ and $\tilde{\delta}$ contribute most significantly to the final tumor volume; however, these contributions are smaller in L[CD] compared to NC[CD].

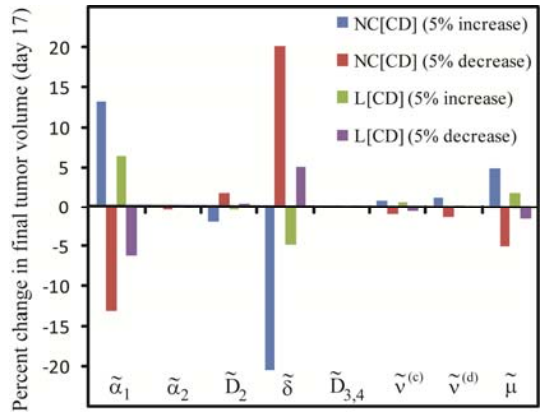

Figure A1: Sensitivity analysis for L[CD] and NC[CD] in Lewis lung carcinoma. Each set comprising four color-coded columns corresponds to the parameter indicated directly below on the x-axis.

To investigate whether the trends obtained from the model for different treatments are robust upon change of these parameters, we performed the numerical simulations for the cases that $\tilde{\alpha}_1$ and $\tilde{\delta}$ are increased by 5% while the rest of the parameters were held fixed. As shown in Fig. A2, these variations do not affect the trends illustrated in Fig. 2a.

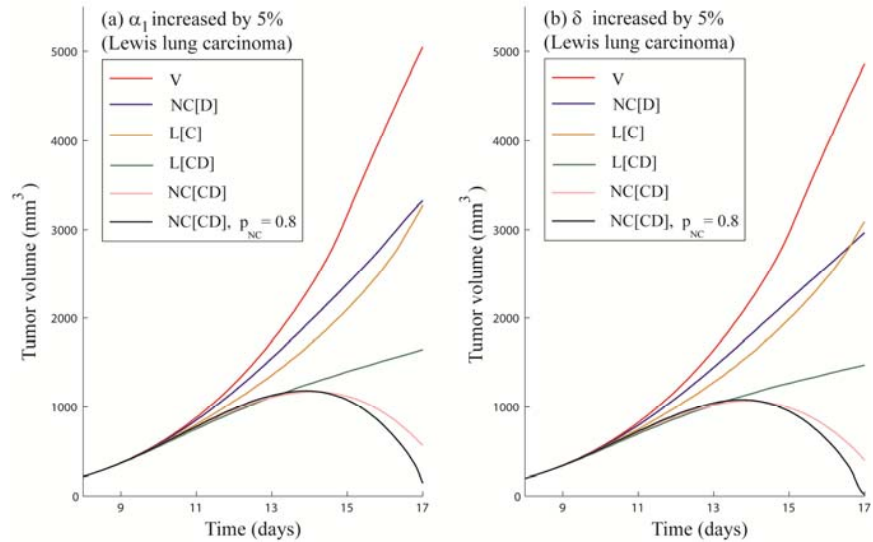

Figure A2: The tumor volume obtained with a 5% increase in (a) $\tilde{\alpha}_1$ and (b) $\tilde{\delta}$ for Lewis lung carcinoma.

**Acknowledgements:** The authors are grateful to G. Powathil for comments on the manuscript. M. Kohandel and S. Sivaloganathan acknowledge financial support by the Natural Sciences and Engineering Research Council of Canada (NSERC) and Canadian Institutes of Health Research (CIHR), C. A. Haselwandter was supported at MIT by an Erwin Schrödinger fellowship of the Austrian Science Fund, and M. Kardar is supported by the NSF Grant No. DMR- 08-03315.

**References**

Batchelor TT, Sorensen AG, di Tomaso E, *et al.* (2007) AZD2171, a pan-VEGF receptor tyrosine kinase inhibitor, normalizes tumor vasculature and alleviates edema in glioblastoma patients, *Cancer Cell* 11:83-95.

Bello L, Carrabba G, Giussani C, *et al* (2001) Low-dose chemotherapy combined with an antiangiogenic drug reduces human glioma growth in vivo, *Cancer Res.* 61:7501-7506.

Bristow RG and Hill RP (2008) Hypoxia and metabolism: Hypoxia, DNA repair and genetic instability, *Nature Reviews Cancer*, 8: 180-192.

de Pillis LG, Radunskaya AE, and Wiseman CL (2005) A validated mathematical model of cell-mediated immune response to tumor growth, *Cancer Res*. 65:7950-7958.

Fenton BM, Paoni SF, and Ding I (2004) Effect of VEGF receptor-2 antibody on vascular function and oxygenation in spontaneous and transplanted tumors, *Radio. Oncol.* 72:221-230.

Franco M, Man S, Chen L, *et al*. (2006) Targeted anti-vascular endothelial growth factor receptor-2 therapy leads to short-term and long-term impairment of vascular function and increase in tumor hypoxia, *Cancer Res*. 66:3639-3648.

Hormigo A, Gutin PH, and Rafii S (2007) Tracking normalization of brain tumor vasculature by magnetic imaging and proangiogenic biomarkers, *Cancer Cell* 11:6-8.

Jain RK (2001) Normalizing tumor vasculature with anti-angiogenic therapy: a new paradigm for combination therapy, *Nature Med.* 7:987-989.

Jain RK (2005) Normalizing tumor vasculature: An emerging concept in anti-angiogenic therapy, *Science* 307:58-62.

Jain RK (2008) Lessons from multidisciplinary translational trials on anti-angiogenic therapy of cancer, *Nature Reviews Cancer* 8:309-316.

Kashiwagi S, Tsukada K, Xu L, *et al*. (2008) Perivascular nitric oxide gradients normalize tumor vasculature, *Nat Med*. 14: 255-7.

Kohandel M, Kardar M, Milosevic M, Sivaloganathan S (2007) Dynamics of tumor growth and combination of anti-angiogenic and cytotoxic therapies, Phys. Med. Biol. **52**, 3665 (2007).

Ma J, Pulfer S, Li S, Chu J, Reed K, and Gallo JM (2001) Pharmacodynamic-mediated reduction of temozolomide tumor concentrations by the angiogenesis inhibitor TNP-470, *Cancer Res.* 61:5491-5498.

Ma J, Waxman DJ (2008) Combination of antiangiogenesis with chemotherapy for more effective cancer treatment, *Molecular Cancer Therapeutics* 7:3670-3684.

Murata R, Nishimura Y, and Hiraoka M (1997) An antiangiogenic agent (TNP-470) inhibited reoxygenation during fractionated radiotherapy of murine mammary carcinoma, *Int. J. Radiat. Oncol. Biol. Phys.* 37:1107-1113.

Perrault SD, Walkey C, Jennings T, *et al*. (2009) Mediating Tumor Targeting Efficiency of Nanoparticles Through Design, *Nano letters* 9: 1909-1915.

Pugh CW and Ratcliffe PJ (2003) Regulation of angiogenesis by hypoxia: role of the HIF system, *Nature Med.* 9: 677-684.

Rofstad EK, Henriksen K, Galappathi K, and Mathiesen B (2003) Antiangiogenic treatment with thrombospondin-1 enhances primary tumor radiation response and prevents growth of dormant pulmonary micrometastases after curative radiation therapy in human melanoma xenografts, *Cancer Res.* 63:4055-4061.

Sengupta S, Eavarone D, Capila I, Zhao G, Watson N, Kiziltepe T, and Sasisekharan R (2005) Temporal targeting of tumour cells and neovasculature with a nanoscale delivery system, *Nature* 436:468-469.

Sengupta S and Sasisekharan R (2007) Exploiting nanotechnology to target cancer, *Br J Cancer* 96:1315-9.

Tong RT, Boucher Y, Kozin SV, Winkler F, Hicklin DJ, and Jain RK (2004) Vascular normalization by vascular endothelial growth factor receptor 2 blockade induces a pressure gradient across the vasculature and improves drug penetration in tumors, *Cancer Res.* 64:3731-3736.

Willett CG, Boucher Y, Tomaso E, *et al.* (2004) Direct evidence that the VEGF-specific antibody bevacizumab has antivascular effects in human rectal cancer, *Nature Med.* 10:145-147.

Winkler F, Kozin SV, Tong RT, *et al.* (2004) Kinetics of vascular normalization by VEGFR2 blockade governs brain tumor response to radiation: Role of oxygenation, angiopoietin-1, and matrix metalloproteinases, *Cancer Cell* 6:553-563.

Yuan F, Leunig M, Huang SK, Berk DA, Papahadjopoulos D, and Jain RK (1994) Microvascular permeability and interstitial penetration of sterically stabilized (stealth) liposomes in a human tumor xenograft, *Cancer Res.* 54:3352-6.

Zips D, Krause M, Hessel F, *et al.* (2003) Experimental study on different combination schedules of VEGF-receptor inhibitor PTK787/ZK222584 and fractionated irradiation, *Anticancer Res.* 23:3869-3876.

**Figure captions**

Figure 1: Confocal micrographs of tissue cross sections harvested from tumor-bearing mice at 24 h post-injection with imaging-nanocells. Mice were injected with nanocells labeled with semiconductor nanocrystals (quantum dots) to monitor the distribution and leakage from the vessels in the tumor and normal tissues. The sections were immunostained for von Willebrand Factor (vWF) to delineate vasculature. Images were captured using a Zeiss LSM510 confocal microscope at 512×512 pixel resolution. The sections were excited with a 488nm laser, and emission was absorbed at FITC (vWF) and Rhodamine (QD) wavelengths. The nanocells were found to be spatially restricted within normal vasculature as seen in the overlap (yellow) of the red and green signal in the merge images but extravasate out from the tumor vasculature at 24h as seen by the predominantly red signal (merge).

Figure 2: Curves for the tumor volume of (a) lung cancer and (b) melanoma obtained with no treatment (V), nanocells containing only doxorubicin (NC[D]), liposomes containing only combretastatin (L[C]), liposomes with combretastatin and doxorubicin (L[CD]), nanocells with combretastatin and doxorubicin (NC[CD]) and nanocells with combretastatin and doxorubicin but with a delayed release of doxorubicin (NC[CD] and $p_{NC}=0.8$). The solid curves are obtained by integrating Eqs. (1.1)–(1.5) in Sec. 2 and the data points are taken from the experiments by Sengupta *et al.* (2005). The same total amount of drugs is released by liposomes and nanocells for the combined therapeutic strategies, which corresponds to double the amount released for NC[D] and L[C] individually.

Figure 3: Mechanisms for the temporal targeting of tumor cells and neovasculature. (a) Nanocells are delivered to the tumor tissue through the neovasculature and rapidly release anti-angiogenic agents. (b) The vascular collapse leads to the trapping of the chemotherapeutic agents within the tumor tissue and thereby prevents re-absorption into the bloodstream. According to our mathematical model this is the principal mechanism responsible for the superior results of the nanocell treatment found by Sengupta *et al.* (2005). (c) The normalization of tumor blood vessels produced by the anti-angiogenic therapy leads to a transient "window of opportunity" (Jain 2005) during which the delivery of nanocells into the tumor tissue is enhanced. (d) Our model suggests that through a judicious timing of the release profiles the interplay between normalization and vascular collapse can be utilized to improve the efficacy of the nanocell treatment.

**Figure 1**

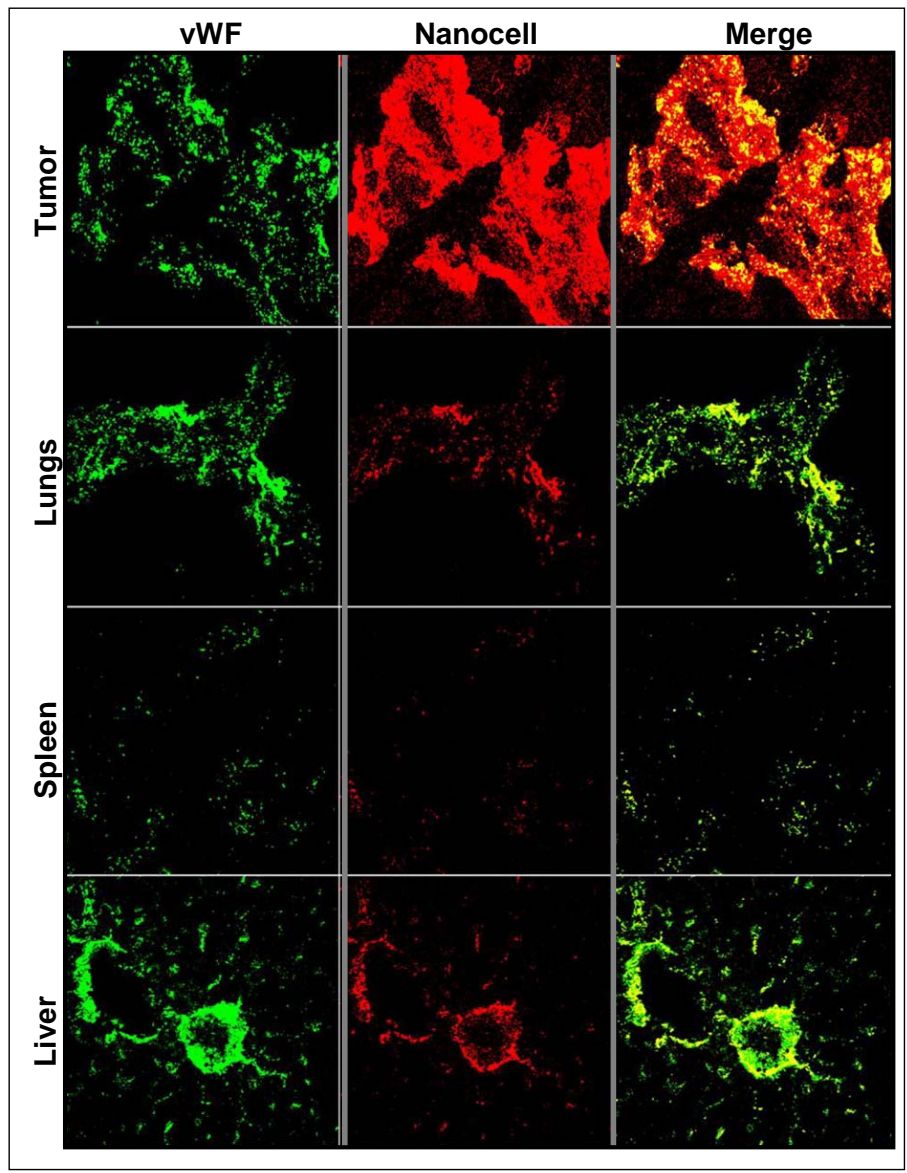

**Figure 2**

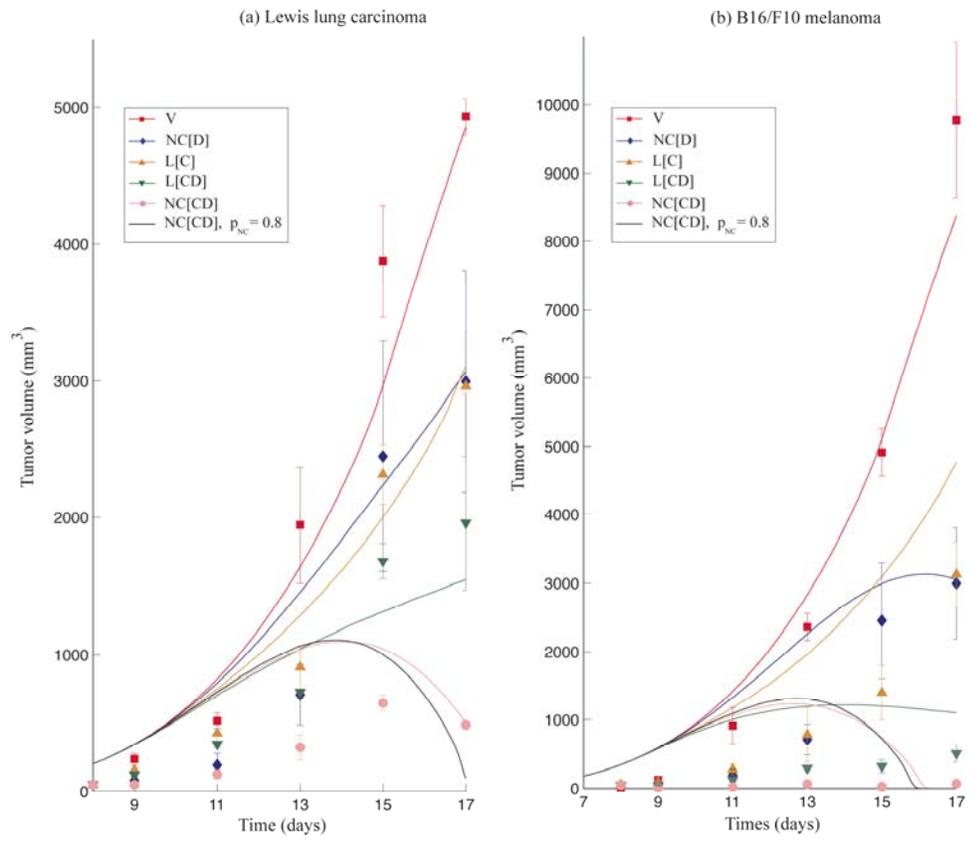

**Figure 3**

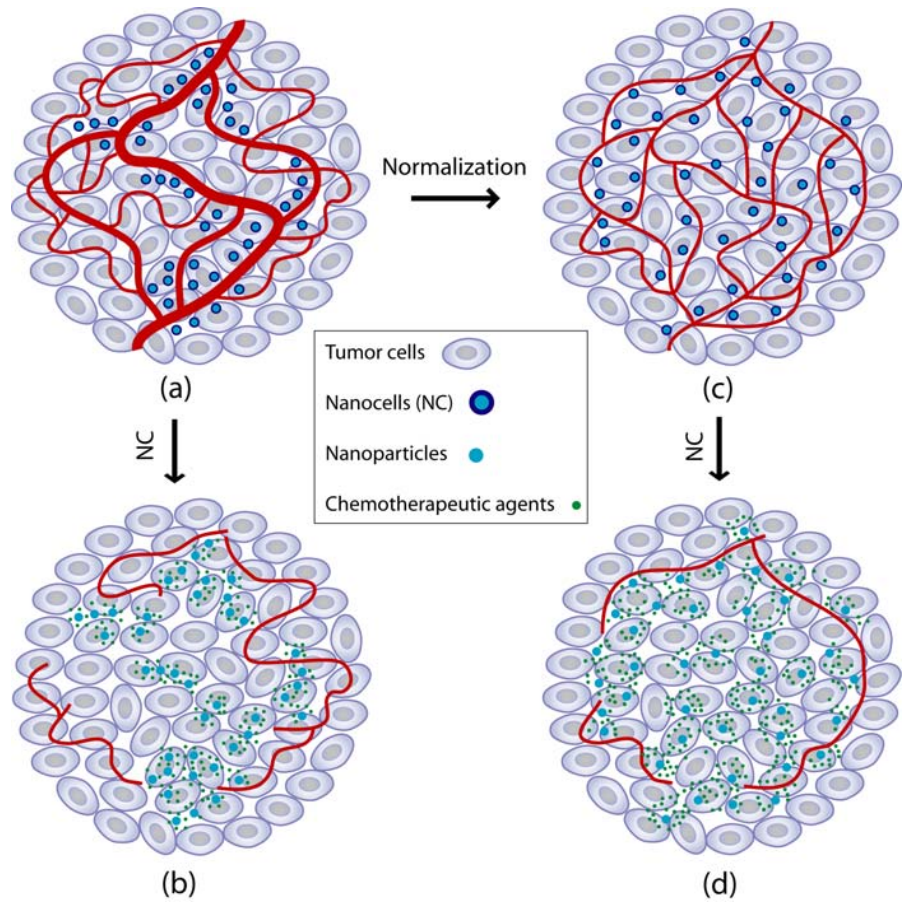